# Topology-dependent thermoluminescence kinetics

Arkadiusz Mandowski

*Institute of Physics, Jan Dlugosz University, ul. Armii Krajowej 13/15, PL-42-200 Czestochowa, Poland,*
*e-mail: a.mandowski@ajd.czest.pl*

Standard one-carrier kinetic models for thermoluminescence (TL) relate to the simple trap model (STM) and the model of localized transitions (LT). This paper presents a review of TL properties in various spatially correlated systems (SCS) which span the two limiting cases. A classification of kinetic models for SCS is proposed. Numerical results are presented for the isolated clusters model (IC) and the interacting 1-D and 3-D systems. For these systems an influence of external electric field on TL is demonstrated. Analytical formulation for the IC model is possible using trap structural functions (TSF's). To calculate the TSF's (analytically or numerically) it is convenient to use their structural and symmetry properties. Analytical equations are presented also for the semi-localized transitions (SLT) model which is a generalized version of LT. It is shown that in some cases the SLT model leads to the cascade detrapping (CD) phenomenon. The CD mechanism produces very narrow TL peaks that are apparently well described by the first order kinetics with very high frequency factors and very high activation energies.

## I. INTRODUCTION

Since the pioneering work by Randal and Wilkins[1] our look at thermoluminescence (TL) models has considerably changed. A miscellany of experimental data - TL kinetic and dose dependence studies, led to subsequent modifications of earlier hypotheses. An inclusion of new trap levels, interactive kinetics and competing centers complicated previous model (for review see[2]). Among the several distinct ways to formulate analytical approximate TL equations, the first order (FO) kinetic approach still enjoys a special status because of its simplicity and strong experimental evidence.

Some additional factors made even more difficult the TL analysis. It was associated with the calling in question the standard assumption of the uniform distribution of traps. The arguments were based on both experimental and theoretical considerations[3]. Using Monte Carlo methods it was shown that spatially correlated TL kinetics cannot be described in terms of the earlier standard models[4,5]. Some peculiarities of TL include an apparent complexity of monoenergetic peaks[6] and the occurrence of 'displacement' peaks due to escaping of charge carriers from their local clusters[7]. Recently, theoretical analysis of this non-standard kinetics has become easier due to analytical formulation of the model[8,9]. In this formulation, topological properties of a material are described by two structural functions irrespective of its thermal history.

This paper presents a review of numerical and analytical studies of TL kinetics in various spatially correlated systems (SCS). These are systems consisting of traps and recombination centres (RC) which are not described by standard TL kinetic models: the model of localized transitions (LT)[10,11] nor by the simple trap model (STM)[2].

## II. ANALYTICAL THEORIES

To classify current TL kinetic models we will consider the two basic models - LT and STM, as the limited cases of TL (Figure 1). Any other types of TL trapping-recombination mechanisms will be attributed to a general SCS model. As a matter of fact, there are lots of topologies belonging to this class. Each of them could be defined by a specific spatial arrangement of traps and RCs with a set of transition probabilities between allowed states. Nevertheless, there are two basic processes permitting smooth topological transition between LT and STM models. The first one is the process of trap (and RC) clustering. The second one is the process of charge carriers' delocalization. These two mechanisms seem to be most important for all SCS. These processes, considered separately, lead to the model of isolated clusters (IC) and the model of semi-localized transitions (SLT), respectively. Fortunately, for these two cases covering a large part of SCS region it was possible to find appropriate analytical models, which will be presented below. All other cases relating to various cluster systems interacting through the conduction band or by direct charge carriers' transitions will be attributed to the interacting clusters (IAC) model. The analysis of IAC systems is possible using Monte Carlo calculations.

### A. The model of isolated clusters (IC)

Let us consider a set of clusters consisting of active trap levels, deep traps and RCs[4]. For the sake of simplicity we will assume that the active charge carriers are electrons. The clusters may have various size (i.e. the number of traps and RCs) and various initial population of charge carriers[12]. TL is produced in the same way as in the LT model - a trapped charge carrier is thermally excited to a local excited level and then it may be retrapped or may recombine with an opposite charge carrier trapped at the RC level. We will assume here, that all trapping and recombination processes occur only within the clusters, i.e. no interaction between clusters is possible. For the simplest case of a single type of active traps, deep traps and RCs it is possible to write the following set of kinetic equations[8]:

$$-\dot{n} = n\nu \exp\left(\frac{-E}{kT(t)}\right) - \Gamma_n(n)n_e, \tag{1a}$$

$$-\dot{h} = \Gamma_h(h)n_e, \tag{1b}$$

$$h = n + n_e + M, \tag{1c}$$

where $E$ stands for the activation energy and $\nu$ is the frequency factor of active traps. $n$, $n_e$ and $h$ denote the total concentrations of electrons trapped in active traps, electrons in the excited levels and holes trapped in RCs. $M$ stands for the concentration of electrons in the thermally disconnected traps (deep traps), i.e. traps that are not emptied during the experiment. $\Gamma_n$ and $\Gamma_h$ denote two functions for trapping and recombination respectively, which depend on the average concentration of charge carriers in the destination level. These functions depend solely on topological properties of the solid, in particular on the distribution of traps and RCs. For this purpose these are called *structural functions*. Their specific properties are discussed in other paper[9]. In every case for sufficiently large clusters the solution of eqs (1) coincide with the STM.

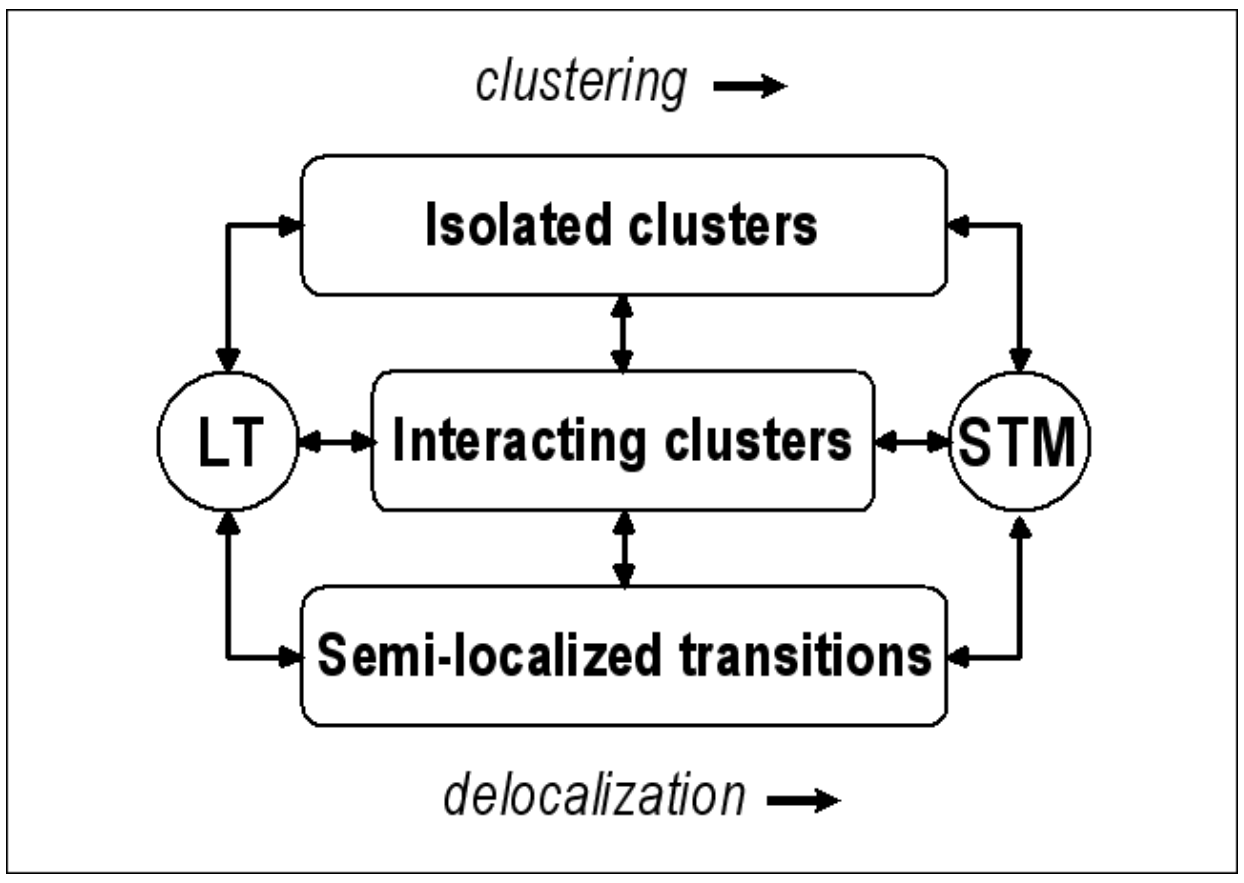

Figure 1. Classification of TL models in spatially correlated systems (SCS). The isolated clusters (IC) model and the model of semi-localized transitions (SLT) are formulated analytically by eqs. (1) and (4), respectively. Two standard models of localized transitions (LT) and the simple trap model (STM) are limiting cases of SCS.

### B. The model of semi-localized transitions (SLT)

The LT model could be completed by considering transitions between the local excited state and the conduction band. Appropriate energy diagram is shown in Figure 2. A similar recombination mechanism, in dose-dependence studies, was recently considered for LiF:Mg,Ti detector[13]. To describe the system analytically we will number all elementary states using the following notation:

$$H_0^0 \equiv \begin{pmatrix}0\\0\\1\end{pmatrix} \quad H_1^0 \equiv \begin{pmatrix}0\\1\\1\end{pmatrix} \quad H_0^1 \equiv \begin{pmatrix}1\\0\\1\end{pmatrix} \quad H_1^1 \equiv \begin{pmatrix}1\\1\\1\end{pmatrix}$$

$$E_0^0 \equiv \begin{pmatrix}0\\0\\0\end{pmatrix} \quad E_1^0 \equiv \begin{pmatrix}0\\1\\0\end{pmatrix} \quad E_0^1 \equiv \begin{pmatrix}1\\0\\0\end{pmatrix} \quad E_1^1 \equiv \begin{pmatrix}1\\1\\0\end{pmatrix} \tag{2}$$

Here the numbers in brackets

$$\begin{pmatrix}\overline{n}_e\\ \overline{n}\\ \overline{h}\end{pmatrix} \tag{3}$$

denote occupation of all states in a single electron-hole pair, i.e. $\overline{n}_e$ is the number of electrons in the local excited level, $\overline{n}$ is the number of electrons in traps and $\overline{h}$ denotes the number of holes in RCs. Therefore, the variables $H_m^n$ and $E_m^n$ denote the concentrations of states with full and empty RCs, respectively.

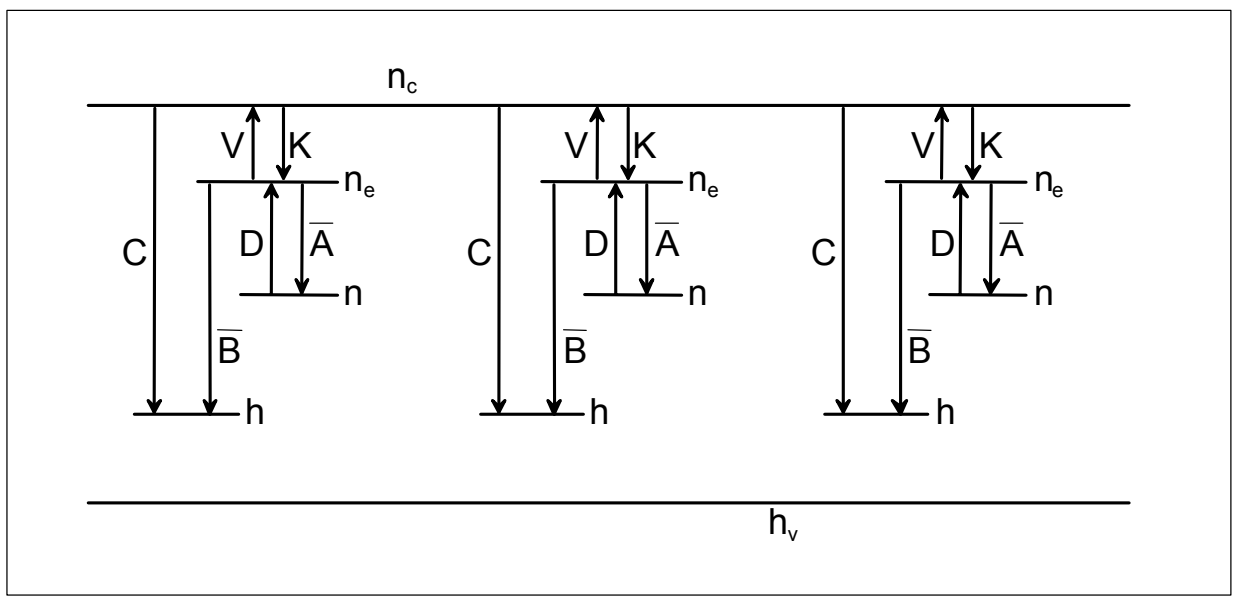

Figure 2. TL kinetics for the SLT model. Intra-pair transitions are denoted by $\overline{A}$ - trapping, $D$ - detrapping and $\overline{B}$ - recombination. Other transitions are denoted as follows: $V$ - excitation to the conduction band, $K$ - capture of a free electron to the excited state of a hole-electron pair and $C$ is the recombination process directly from the conduction band.

Initial excitation generates only $H_1^0$ states. For simplicity we will consider the SLT (Figure 2) without $K$ transitions, i.e. $K = 0$. This assumption eliminates the states $H_1^1$ and $E_1^1$, then allows to write the following set of differential equations:

$$\dot{H}_1^0 = -\left(D + Cn_c\right)H_1^0 + \overline{A}H_0^1 \tag{4a}$$

$$\dot{H}_0^1 = DH_1^0 - \left(\overline{A} + \overline{B} + V + Cn_c\right)H_0^1 \tag{4b}$$

$$\dot{H}_0^0 = VH_0^1 - Cn_cH_0^0 \tag{4c}$$

$$\dot{E}_1^0 = Cn_cH_1^0 - DE_1^0 + \overline{A}E_0^1 \tag{4d}$$

$$\dot{E}_0^1 = Cn_cH_0^1 + DE_1^0 - \left(\overline{A} + V\right)E_0^1 \tag{4e}$$

$$\dot{E}_0^0 = \overline{B}H_0^1 + Cn_cH_0^0 + VE_0^1 \tag{4f}$$

$$\dot{n}_c = -Cn_c(H_1^0 + H_0^1 + H_0^0) + V(H_0^1 + E_0^1) \tag{4g}$$

Here $\overline{B}$ and $C$ denote coefficients for direct (intra-pair) and band-to-RC recombination, respectively. $\overline{A}$ denotes coefficient for trapping. The excitation coefficients $D$ and $V$ are thermally activated

$$D(t) = \nu \exp\left(\frac{-E}{kT(t)}\right) \tag{5}$$

$$V(t) = \nu_V \exp\left(\frac{-E_V}{kT(t)}\right) \tag{6}$$

where $E$, $E_V$, $\nu$ and $\nu_V$ are constants. Nonetheless, as the states $H_m^n$ and $E_m^n$ are virtually different, it is reasonable to allow different values of the coefficients (5,6) for these two classes.

## III. BASIC PROPERTIES OF SPATIALLY CORRELATED SYSTEMS

### A. The model of isolated clusters (IC)

Numerical solutions for the IC model could be obtained by calculating two Γ-functions[9] and then, solving the set of eqs (1). Another possibility is a direct Monte Carlo simulation method using elementary transitions for each cluster separately ($\Im_D$, $\Im_T$ and $\Im_R$ for detrapping, trapping and recombination, respectively):

$$\Im_D(t) = \nu \exp\left[\frac{-E}{kT(t)}\right] \quad (7a)$$

$$\Im_T(t) = \overline{A}\left[\overline{N} - \overline{n}(t)\right] \quad (7b)$$

$$\Im_R(t) = \overline{B}\overline{h}(t) \quad (7c)$$

Here $\overline{N}$ denotes the number of trap levels in each cluster. Detailed method of the simulation as well as the scaling properties defining relation between the microscopic (used here) and the macroscopic parameters (used e.g. in the STM) were given in some previous papers[4-6].

It was found that major discrepancies between TL glow curves calculated for IC model and the standard TL models (LT and STM) occur for rather high retrapping coefficients ($r \equiv \overline{A}/\overline{B} \geq 1$) and small cluster systems ($1 < \overline{N} < 10$)[4,14] (Figure 3). Usually, these curves could not be described in terms of first/second nor general order kinetics[6]. However, an excellent agreement was obtained while fitting TL with a sum of $p$ FO peaks where the number of peaks is the same as the number of initially trapped charge carriers in a single cluster $\overline{n}_0$ [14]. Therefore, the number of peaks provides an information about the cluster's size. Calculated activation energies and frequency factors are very close in this case to the real values (cf. Figure 4). A similar conclusion may be done with respect to partially filled clusters. As a general rule the small-cluster contribution to TL may be expressed as a sum of several FO peaks.

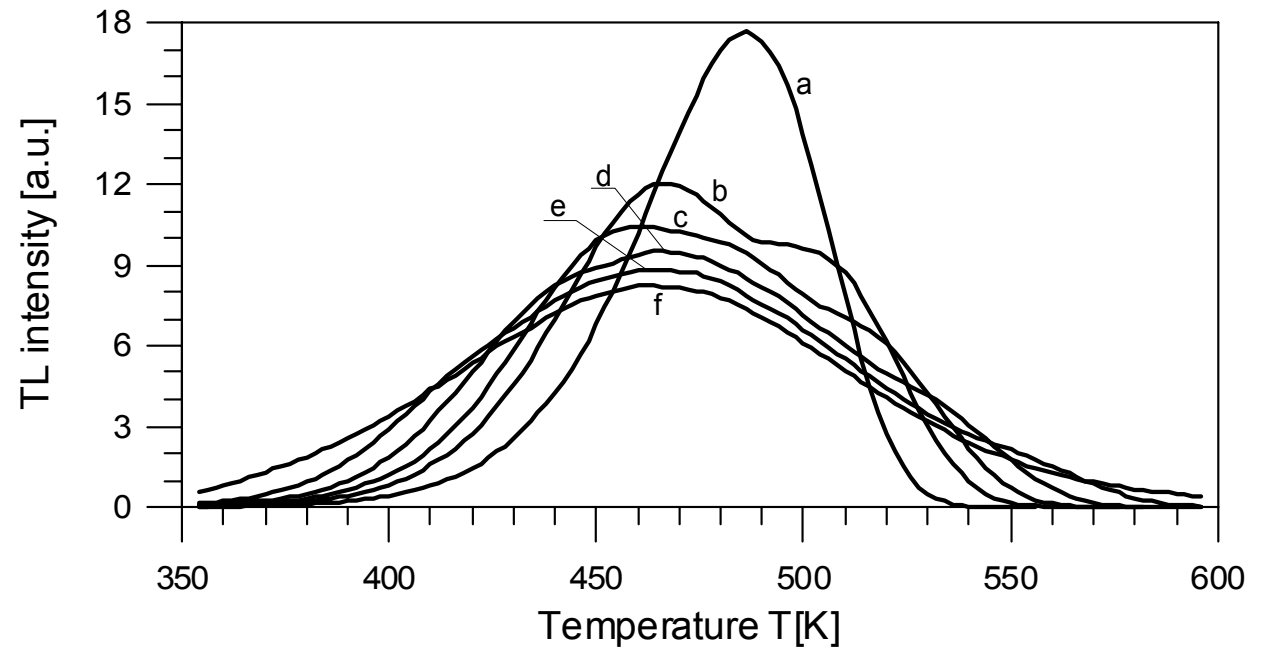

Figure 3. The dependence of the shape of TL spectra on the number of carriers in a single cluster of traps $\overline{n}_0$. The curves were calculated for a fully populated ($\eta_0 \equiv \overline{n}_0/\overline{N} = 1$) IC system characterized by $r = 100$ with no 'deep traps'. The calculated curves correspond to the following cases: a) $\overline{n}_0 = 1$ (LT model), b) $\overline{n}_0 = 2$, c) $\overline{n}_0 = 3$, d) $\overline{n}_0 = 5$, e) $\overline{n}_0 = 10$ and f) $\overline{n}_0 = 100$. For $\overline{n}_0 \geq 100$ all TL curves coincide with the STM model.

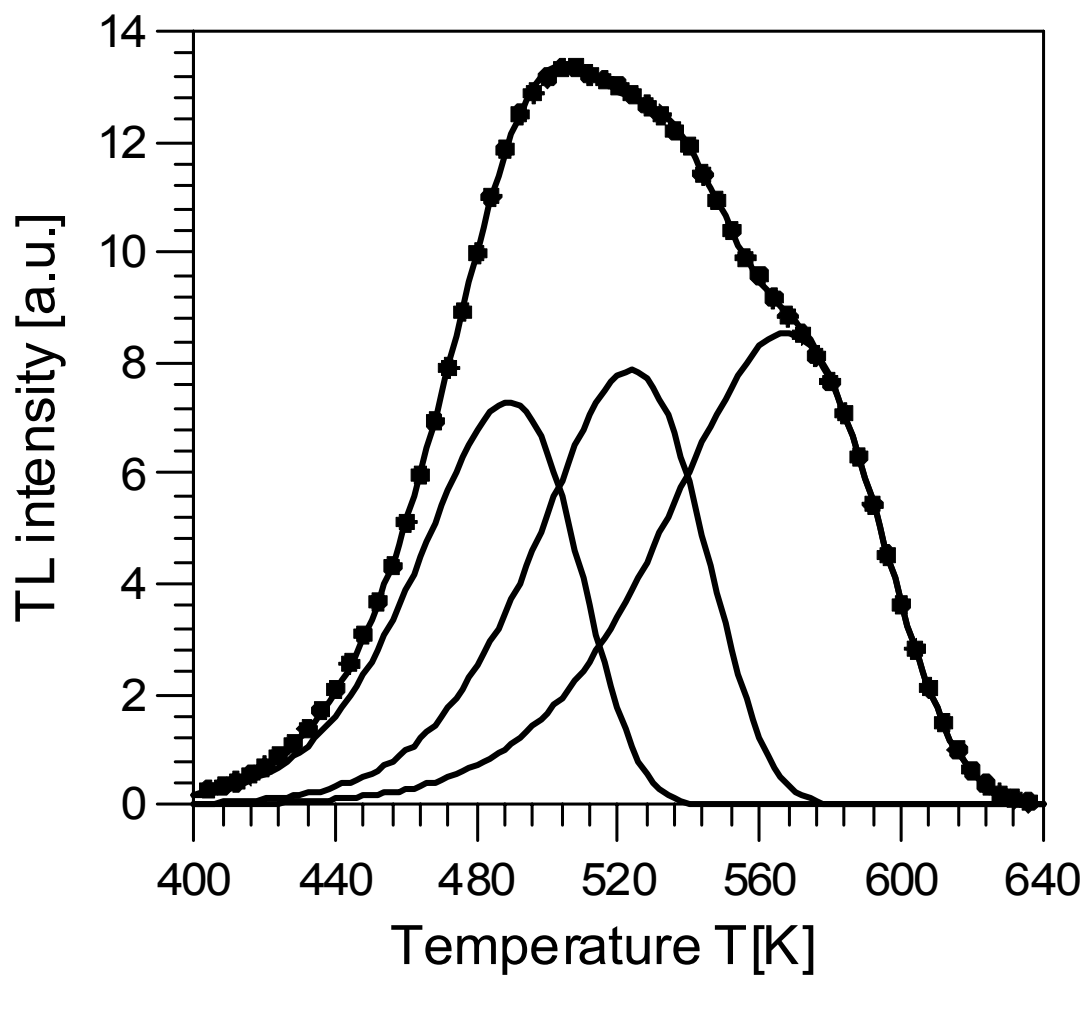

Figure 4. Results of glow curve deconvolution performed for a single 'spatially correlated' TL peak calculated for $E = 0.9\,\text{eV}$, $\overline{n}_0 = 3$ and $r = 10^3$. The curve can be perfectly deconvoluted for three FO peaks having 'energies' $E_1 = 0.893\,\text{eV}$, $E_2 = 0.911\,\text{eV}$ and $E_3 = 0.932\,\text{eV}$.

**B. Multi-level IC systems**

The same method of Monte Carlo simulations (7) may be applied to a system of clusters containing several trap levels and RCs[14]. Typical results for two-level system characterized by close activation energies $E_1 = 0.9\,\text{eV}$ and $E_2 = 1.0\,\text{eV}$ and the same frequency factors $\nu_1 = \nu_2 = 10^{10}\,\text{s}^{-1}$ are shown in Figure 5. Due to high retrapping coefficient "classical" TL glow curve (i.e. calculated in the framework of the simple model - curve *d*) does not reveal its complex structure. Along with decreasing the number of charge carriers in a single group TL curve changes giving for purely localized transitions two well separated peaks. Thus, strong spatial correlation helps identify closely positioned energy levels.

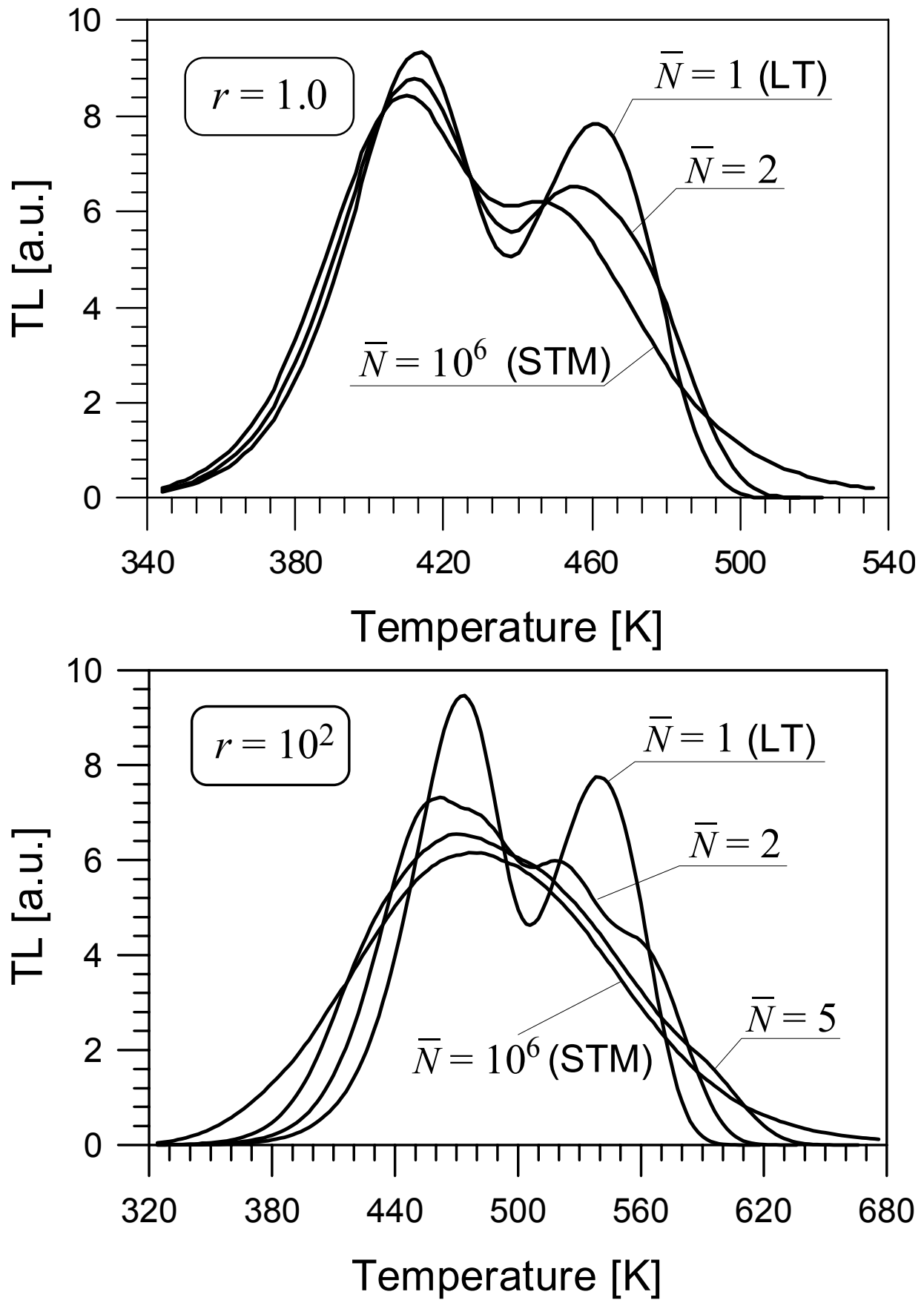

Figure 5. The dependence of the shape of TL on the cluster size $\overline{N}$ in a two-level system with two different activation energies: $E_1 = 0.9\,\text{eV}$ and $E_1 = 1.0\,\text{eV}$ calculated for two retrapping coefficients: $r = 1$ and $r = 100$. Other parameters: $\nu_1 = \nu_2 = 10^{10}\,\text{s}^{-1}$, $\eta_{01} = \eta_{02} = 1$, and the heating rate $\beta = 1\,\text{K/s}$.

### C. Modelling of SCS with transport properties. The influence of electric field.

When transitions of carriers between neighbouring clusters could not be neglected the whole system (i.e. all groups of traps) has to be considered simultaneously during the Monte Carlo simulations. It has to be done, e.g. while considering the effect of an external electric field on trapping and recombination kinetics. To deal with this problem one has to assume a certain spatial distribution of clusters. Two basic classes were considered - one-dimensional chains[7,15] and a cubic network of cells (clusters)[5]. In all cases periodic boundary conditions were applied to avoid difficulties with too small size of the network. Along with the three types of allowed transitions (7a-c) a carrier in the excited state has a chance to jump over the energy barrier to an adjacent group. The transition probability $A_{tr}$ is thermally activated

$$A_{tr} = \nu_{tr} \exp\left(\frac{-E_{tr}}{kT}\right) \qquad (8)$$

where $E_{tr}$ and $\nu_{tr}$ are constants. The calculations show that even for a simple system with a monoenergetic trap level for electrons, TL curves may exhibit complex structure. An additional high-temperature peak (the *displacement peak*) can be observed due to transitions of charge carriers between groups of traps. For very high transition probabilities $A_{tr}$ as compared to the recombination probability (7c) the displacement peak increases and it mingles with the main TL peak. Figure 6 illustrates the influence of the electric field on TL curves for a one-dimensional system. The strength of the field is characterised by the traps' potential lowering $\Delta E_{field}$. For low electric fields (the upper diagram A), only the displacement peak changes its height and position. In a more intense field (the bottom diagram B), the two peaks merge into a single one. Its shape depends on the field strength. Results obtained for a cubic network are qualitatively the same. Only for very low values of the external field the displacement peak is slightly bigger and shifted to higher temperatures.

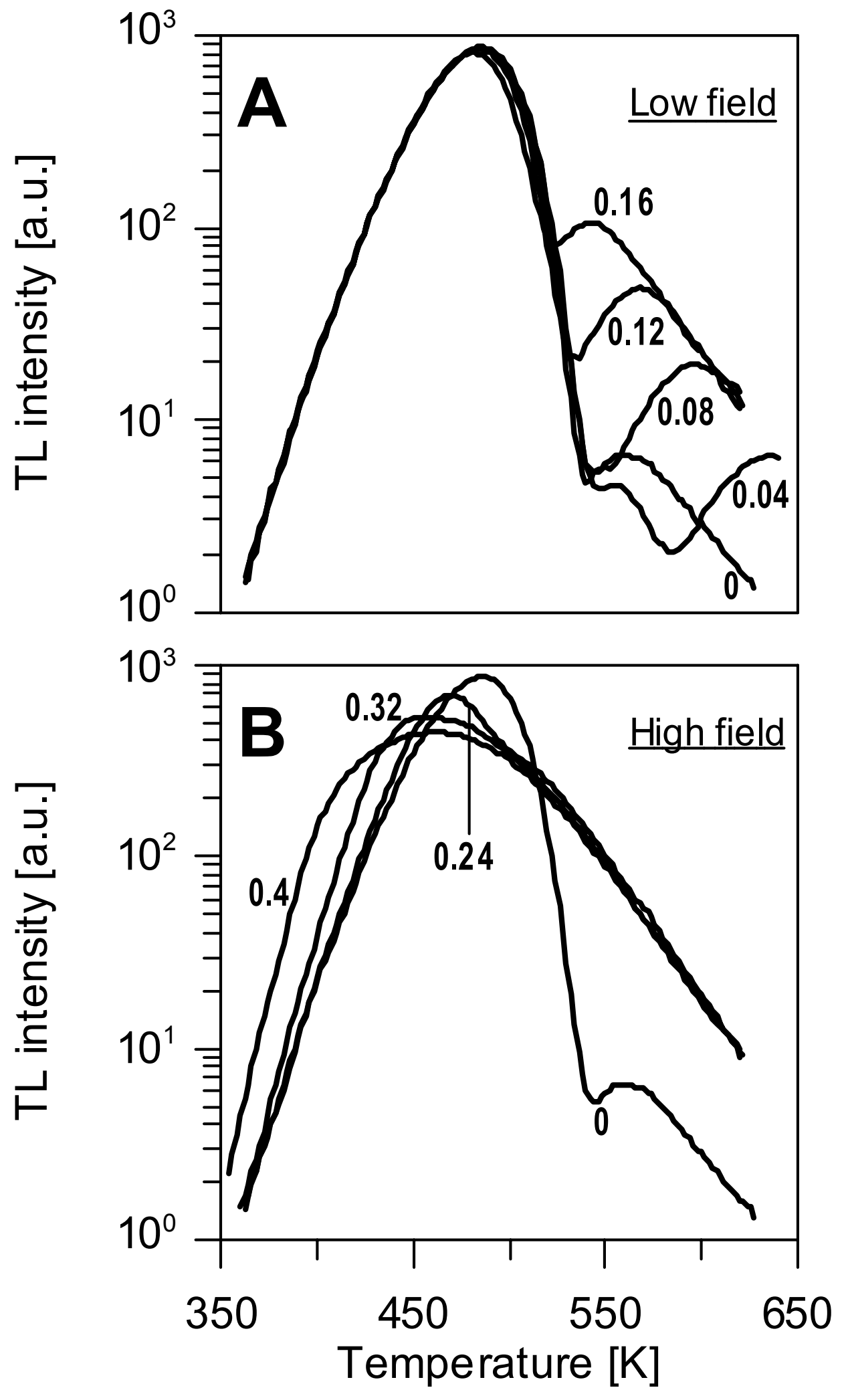

Figure 6. TL curves calculated for various values of the external electric field: $\Delta E_{field}$ shown on the diagrams. Transition rate parameters: $E_{tr} = 0.8\,\mathrm{eV}$ and $\nu_{tr} = 10^9\,\mathrm{s}^{-1}$. Trap parameters: $E = 0.9\,\mathrm{eV}$, $\nu = 10^{10}\,\mathrm{s}^{-1}$ and $r = 100$. The 'zero field curve' is shown on two diagrams for reference.

### D. SLT and the cascade detrapping (CD) effect

In some cases, when particular charge carrier transitions (7a-c,8) differ several orders of magnitude, the time required for Monte Carlo calculations is too high for computations even on fastest computers. Analytical formulation of the SLT model eliminates the necessity for performing Monte Carlo simulations. Here the solution may be obtained by solving the set of nonlinear equations (4). There are two possible radiative transitions - $\mathcal{L}_{\bar{B}}$ . and $\mathcal{L}_C$ . The first one relates to direct recombination within the hole-electron pair. The second one denotes recombination from the conduction band. The luminescence is defined by:

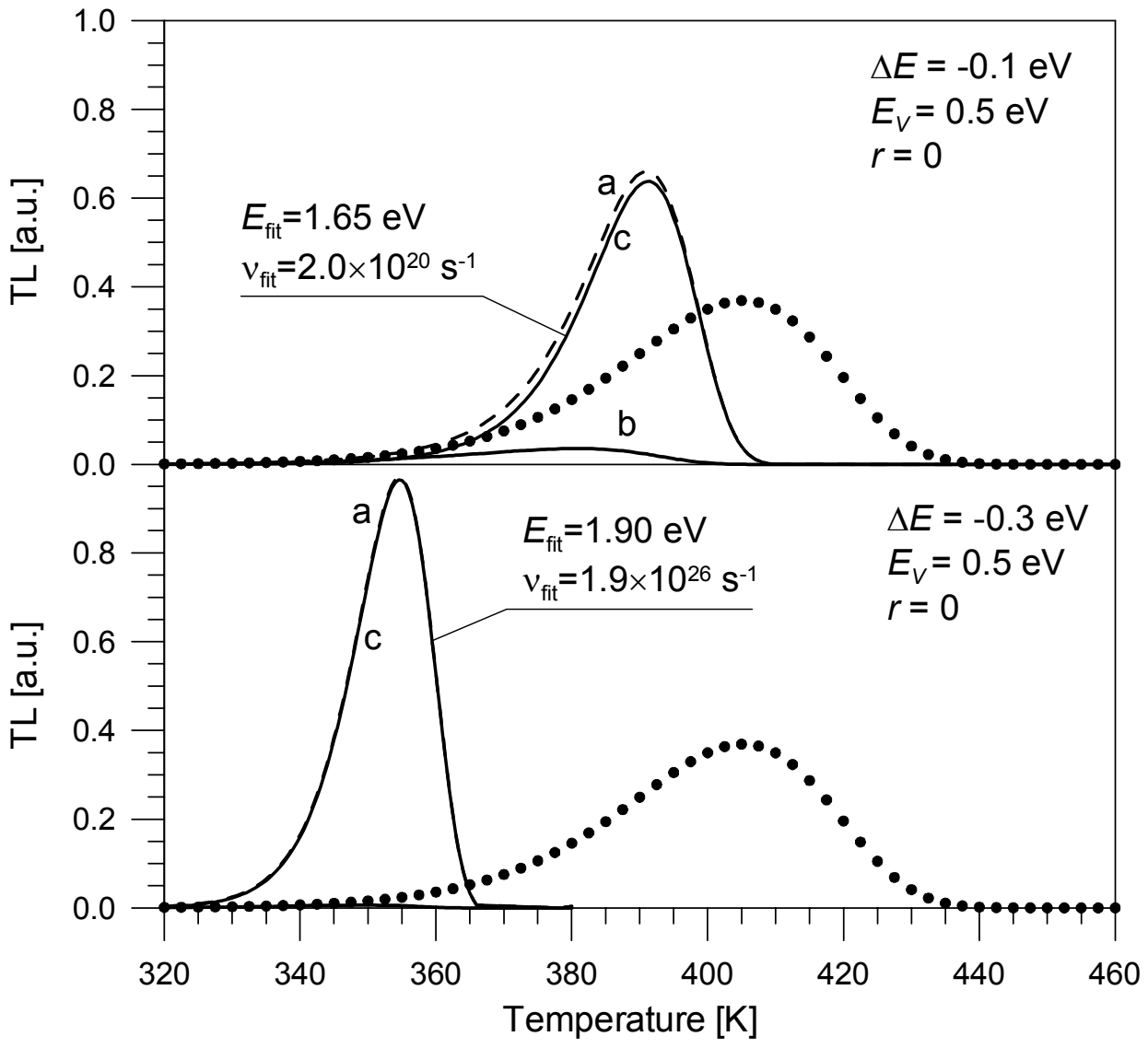

Figure 7. Illustration of the cascade detrapping phenomenon. TL glow curves calculated in the framework of SLT model with variable energy. The activation energy for $H_1^0$ states is $E_1 = 0.9\,\mathrm{eV}$. For $E_1^0$ the energy changes to $E_2 = E_1 + \Delta E$, where $\Delta E$ is shown on the diagrams. All other trap parameters remain unchanged: $E_V = 0.5\,\mathrm{eV}$, $\nu = \nu_V = 10^{10}\,\mathrm{s}^{-1}$, $r = 0$ and $\beta = 1\,\mathrm{K/s}$. All curves *b* correspond to direct transitions $\mathcal{L}_{\bar{B}}$ (9), curves *c* correspond to $\mathcal{L}_C$ and *a* correspond to total intensities $\mathcal{L}_{\text{tot}} = \mathcal{L}_{\bar{B}} + \mathcal{L}_C$. Black dots (•) denote related solution for the LT model. $\mathcal{L}_C$ intensities were fitted to the FO model. Fitted values $E_{\text{fit}}$ and $\nu_{\text{fit}}$ are shown on the diagrams.

$$\mathcal{L}_{\overline{B}} = \overline{B}H_0^1 \quad (9)$$

$$\mathcal{L}_C = Cn_c(H_1^0 + H_0^1 + H_0^0) \quad (10)$$

Below, we will limit ourselves to the description of the cascade detrapping (CD) phenomenon which can provide an explanation for very high values of frequency factors and activation energies observed in many TL detectors (e.g. LiF:Mg,Ti and LiF:Mg,Cu,P[2]). The CD mechanism occurs when the activation energy for detrapping is higher for a cell with full RC (i.e. $H_1^0$) than for a cell with empty RC ($E_1^0$). Let us denote the activation energies by $E_1$ and $E_2$, respectively. The change of activation energy will be denoted by $\Delta E = E_2 - E_1$. Typical results for two values $\Delta E = -0.1\text{eV}$ and $\Delta E = -0.3\text{eV}$ are presented in Figure 7. The peak $\mathcal{L}_C$ is very narrow and shifted to lower temperatures as compared with the standard solution for appropriate LT model. This peak may be accurately fitted by the FO equation. Fitted activation energies and frequency factors are very high. The TL curve in Figure 7 was calculated for $E = 0.9\text{eV}$ and $\nu = 10^{10}\,\text{s}^{-1}$ whereas the fitted values are $E_{\text{fit}} = 1.65\text{eV}$ with $\nu_{\text{fit}} = 2.0\times10^{20}\,\text{s}^{-1}$ (for $\Delta E = -0.1\text{eV}$) and $E_{\text{fit}} = 1.90\text{eV}$ with $\nu_{\text{fit}} = 1.9\times10^{26}\,\text{s}^{-1}$ (for $\Delta E = -0.3\text{eV}$). It should be noted, that similar conclusions were obtained by Piters and Bos[16] who considered a defect-interaction model. Its relation to the SLT will be considered in a separate paper.

## VI. CONCLUSIONS

For the last four decades, a majority of TL glow curve analyses were restricted to STM and LT models, typically in their simplified versions. Actually, there are no physical reasons (except mathematical ones) why the TL kinetics should be limited to the two classical cases. The model for SCS, presented here, offers a smooth topological transition between LT and STM. To classify particular cases (i.e. sub-models) we distinguished two basic processes responsible for the topological transition. These are: clustering of traps and delocalization of active charge carriers. Starting from the LT model, these two processes, considered separately, lead to analytically formulated sub-models - IC and SLT. Analytical formulation appreciably facilitates examination of TL kinetic properties in these systems. The most complex IAC systems may be studied by means of the Monte Carlo method.

Some properties of TL in SCS are unique and far different from that of what can be learned from LT and STM models. Typical examples include e.g. the dependence of TL on the external electric field in 1-D and 3-D systems. Furthermore, it was shown that small-cluster contribution to TL takes the form of a sum of several FO peaks. Numerical analysis of SLT model revealed new TL effect that we call the *cascade detrapping* phenomenon. The CD mechanism leads to narrow TL peaks well described by FO kinetic equation with very high frequency factors and activation energies. These properties could explain some unusual kinetic features observed in many TL detectors. For a direct proof, however, still some theoretical and experimental work is necessary. Undoubtedly, one of the main goals of TL modelling is the unification of kinetic and dose-dependence models. The SCS kinetic model seems to be a good platform for these efforts.

## V. ACKNOWLEDGEMENTS

This work was supported by the Polish State Committee for Scientific Research grant no. 3T10C01326

---

1. Randall, J. T. and Wilkins, M. H. F. *Phosphorescence and electron traps: I. The study of trap distributions.* Proc. R. Soc. A **184**, 366–389 (1945).
2. Chen, R. and McKeever, S.W.S. *Theory of thermoluminescence and related phenomena*, (Singapore - World Scientific). ISBN 9810222955. (1997).
3. Townsend, P. D. and Rowlands, A. P. *Extended defect models for thermoluminescence.* Radiat. Prot. Dosim. **84**(1-4), 7-12 (1999).
4. Mandowski, A. and Świątek, J. *Monte Carlo simulation of thermally stimulated relaxation kinetics of carriers trapping in microcrystalline and two-dimensional solids.* Phil. Magazine B, **65** 729-732 (1992).
5. Mandowski, A. and Świątek, J. *Trapping and recombination properties due to trap clustering.* J. Phys. III (France) **7**, 2275-2280 (1997).
6. Mandowski, A. and Świątek, J. *On the influence of spatial correlation on the kinetic order of TL.* Radiat. Prot. Dosim. **65**(1-4), 25-28 (1996).
7. Mandowski, A. *One - dimensional thermoluminescence kinetics*, Radiat. Measurements **33**, 747-751 (2001).
8. Mandowski, A. *The theory of thermoluminescence with an arbitrary spatial distribution of traps.* Radiat. Prot. Dosim. **100**(1-4), 115-118 (2002).
9. Mandowski, A. *Calculation and properties of trap structural functions for various spatially correlated systems.* Radiat. Prot. Dosim. (submitted to this issue)
10. Halperin, A. and Braner, A. A. *Evaluation of thermal activation energies from glow curves.* Phys. Rev. **117**, 408-415. (1960).
11. Land, P. L. *Equations for thermoluminescence and thermally stimulated current as derived from simple models*. J. Phys. Chem. Solids **30**, 1693-1708 (1969).
12. Mandowski, A. *On the analysis of TL glow curves in spatially correlated systems*. Radiat. Prot. Dosim. **84**(1-4) 21-24 (1998).
13. Nail, I., Horowitz, Y. S., Oster, L. and Biderman, S. *The unified interaction model applied to LiF:Mg,Ti (TLD-100): properties if the luminescent and competitive centres during sensization.* Radiat. Prot. Dosim. **102**(4) 295-304 (2002).
14. Mandowski, A. and Świątek, J. 1998 *Thermoluminescence and trap assemblies - results of Monte Carlo calculations*. Radiat. Measurements **29**(3/4), 415-419 (1998).
15. Mandowski A., *Modelling of charge carriers' transport and trapping in one-dimensional structures during thermal stimulation*, J. Electrostatics **51-51**, 585-589 (2001).
16. Piters, T. M. and Bos A. J. J. A model for the influence of defect interactions during heating on thermoluminescence in LiF:Mg,Ti (TLD-100). J. Phys. D: Appl. Phys. **26**, 2255-2265 (1993).